\documentclass[preprint,12pt]{elsarticle}

\usepackage{epsfig}

\usepackage{amssymb}
\usepackage{amsmath}

\begin{document}

\begin{frontmatter}



\title{Different Types of Synchronization in Coupled Network Based Chaotic Circuits}

\author[rvt1]{K.~Srinivasan \corref{cor1}}
\author[rvt2]{V.K~Chandrasekar}
\author[rt]{R.~Gladwin Pradeep}
\author[els]{K.~Murali}
\ead{kmurali@annauniv.edu}
\author[rvt]{M.~Lakshmanan }
\ead{lakshman@cnld.bdu.ac.in}

\cortext[cor1]{Corresponding author}
\address[rvt1]{Department of Physics, Nehru Memorial College, Puthanampatti, Tiruchirapalli 621 007, India}
\address[rvt2]{Centre for Nonlinear Science \& Engineering, School of Electrical \& Electronics Engineering, SASTRA University, Thanjavur 613 401, India}
\address[rt]{Department of Physics, KCG College of Technology, Chennai 600 097, India}
\address[els]{Department of Physics, Anna University, Chennai 600 025, India}
\address[rvt]{Centre for Nonlinear Dynamics, Department of Physics,
Bharathidasan University, Tiruchirapalli 620 024, India}

\begin{abstract}
We propose a simple and new unified method to achieve lag, complete and anticipatory synchronizations in coupled nonlinear systems.  It can be considered as an alternative to the subsystem and intentional parameter mismatch methods. This novel method is illustrated in a unidirectionally coupled RC phase shift network based Chua's circuit.  Employing feedback coupling, different types of chaos synchronization are observed experimentally and numerically in coupled identical chaotic oscillators {\emph{without using time delay}}.  With a simple switch in the experimental set up we observe different kinds of synchronization.  We also analyze the coupled system with numerical simulations. 

\end{abstract}


\end{frontmatter}


\section{\label{sec1}Introduction}
Synchronization is a fundamental and ubiquitous phenomenon, first discovered in coupled pendula by Huygens~\cite{huygens}.
Chaos synchronization properties of uni- and bidirectionally coupled chaotic systems have attracted the attention of many researchers due to their potential applications in a variety of fields~\cite{pecora1,cuomo,lakmurbook,chua,parlitz,taherion,rosenblum2,buscarino,chen,agrawal,yassen}.  Different synchronization states have been studied in the literature, including complete or identical synchronization (CS)~\cite{pecora1,lakmurbook,murali2}, in-phase synchronization(PS)~\cite{rosenblum1,srini3,dana}, antiphase synchronization (APS)~\cite{liu,srini1,ambika2}, lag synchronization (LS)~\cite{rosenblum2,srini1}, anticipatory synchronization (AS)~\cite{srini1,ambika1,voss1,voss2}, and generalized synchronization (GS)~\cite{kocarev}. In the case of lag synchronization ~\cite{rosenblum2,srini1} the response system trails behind the drive system, while during  anticipatory synchronization~\cite{srini1,ambika1,voss1,voss2} the response system leads the drive system in time. Such time-shifted synchronizations have been observed in lasers~\cite{wu}, neuronal models~\cite{pikovsky}, and electronic circuits~\cite{srini1} and offer intriguing possibilities for possible technological applications~\cite{cuomo,pecora2,murali3,murali1,grosu}.

Using an explicit time delay or a memory unit, both lag and anticipatory synchronizations in unidirectionally coupled oscillators can be observed.  Lag synchronization can be obtained by coupling the response system to a past state of the drive, whereas anticipatory synchronization can be obtained through a feedback mechanism where the current state of the drive system is coupled to the past state of the response system~\cite{pikovsky}.  In either case, an explicit time delay appears in the coupling.  

An approximate lag synchronization in mutually coupled chaotic oscillator can be observed by having a parameter mismatch~\cite{corron1,corron2,corron3}.    In particular, intermittent and continuous lag synchronizations have been observed as intermittent steps in a route from phase to complete synchronization as the coupling strength is increased~\cite{pikovsky}.  In general, both lag and anticipatory synchronizations in unidirectionally coupled chaotic oscillators can be achieved using a specific intentional parameter mismatch between the drive and response systems~\cite{corron1}.  This type of lag and anticipatory synchronizations has important technological implications in engineered systems.

In this paper we propose an alternative approach to the subsystem approach and the intentional parameter mismatch method and achieve different types of synchronization in a network of unidirectionally coupled identical Chua's circuit.  Chua's circuit and its variance are well known chaotic circuits which exhibit a wide variety of nonlinear dynamical phenomena such as bifurcations and chaos~\cite{lakmurbook,ogo,chenueta,madan,chukom,kenn1992}.
Recently, three of the present authors along with others have constructed a new phase shift network based Chua's circuit~\cite{srini2}, which exhibits period-doubling bifurcation route to chaotic attractor.  In this manuscript, we study the occurrence of different types of chaos synchronization in unidirectionally coupled system of two identical RC phase shift network based Chua's circuits~\cite{srini2}.  We first demonstrate experimentally and then verify them by numerical methods (which is needed to confirm the similarity functions as explained in the text) and obtain lag, complete and anticipatory synchronizations.  The method involves switching the connection in the drive and response circuits which in turn produces the above types of synchronization. In short, this mechanism is a very simple and elegant way of achieving lag, complete and anticipatory synchronizations in unidirectionally coupled chaotic oscillators.      

The organization of this paper is as follows.  In Sec. 2, the circuit realization of RC phase shift network based Chua's circuit is discussed.  The nature of synchronizations of this coupled Chua's circuit is studied both experimentally and numerically for different coupling configurations.  We summarize the results in Sec. 3.

\section{\label{sec:2}Coupled RC phase shift network based Chua's circuit}

\begin{figure}
\centering
\includegraphics[width=1.0\columnwidth]{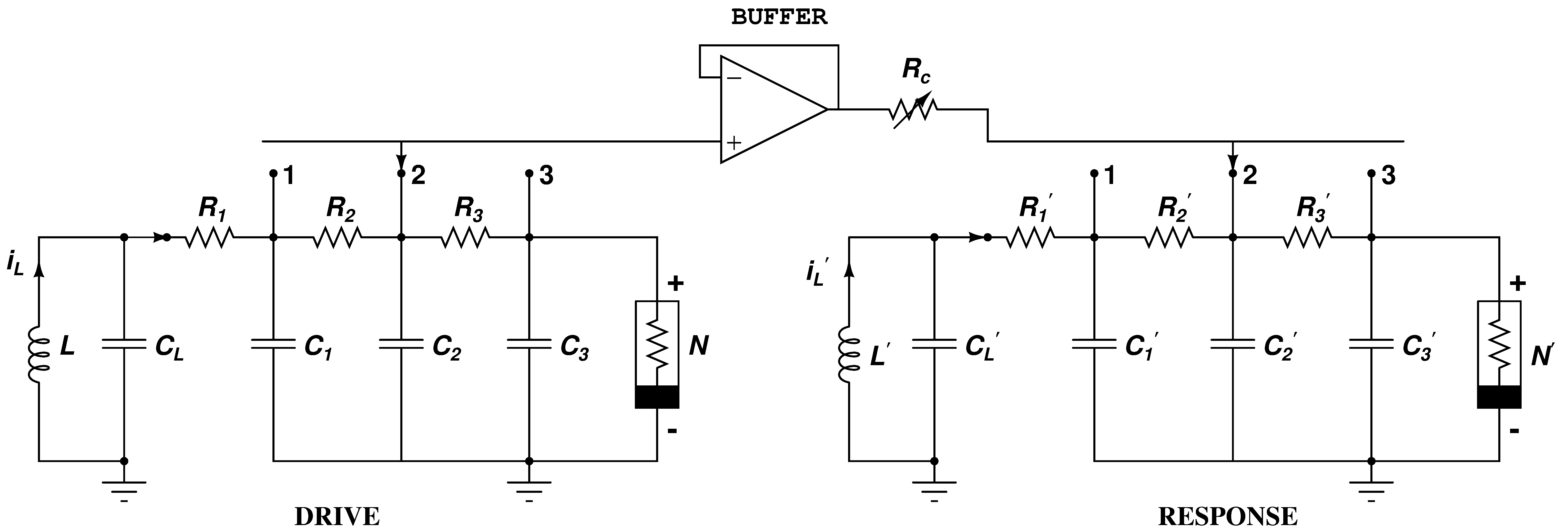}
\caption{\label{chuam_error} Circuit realization of unidirectional feedback coupled phase shift network based Chua's circuit.}
\end{figure}

In the following we study the dynamics of a system of two coupled identical RC phase shift network based modified Chua's circuits which is shown in Fig.~\ref{chuam_error}.  In this circuit, $v_1$, $v_2$, $v_3$, $v_{LC}$ and $i_L$ are the voltages across the capacitors $C_1$, $C_2$, $C_3$ and $C_L$ and the current through the inductor $L$, respectively.  The drive system is connected to the response system through a buffer and a coupling resistance ($R_c$). The buffer allows the current to flow from drive to response circuit only. Using the unidirectional feedback coupling approach, we demonstrate three types of chaos synchronization, namely complete, lag and anticipatory synchronizations in the proposed modified coupled Chua's circuit both experimentally and numerically without the introduction of any time-delay or parameter mismatch.  We also note that the motivation for carrying out numerical studies here is also to confirm the nature of the similarity functions in the following.  By simply connecting (switching) the response system and the drive system through any of the three terminals $1$, $2$ and $3$ of the drive and response systems (shown in Fig.~\ref{chuam_error}), we observe the three types of synchronization.  The reason for introducing different couplings of state variables $v_1$ , $v_2$ and $v_3$ is to observe all the three types of synchronizations, namely lag, identical and anticipatory synchronizations, in a simple manner. This is achieved by exploiting the finite phase-shift that is being introduced by the individual RC network elements.  We use the notation $(m,n)$ to denote that the $m^{\mbox{th}}$ terminal of the drive system is connected to the $n^{\mbox{th}}$ terminal of the response system.  The details are as follows.

\subsection{Drive system connected to the terminal $2$ of the response system  }
In this subsection we study the dynamics of the coupled RC phase shift network based Chua's circuit (Fig.~\ref{chuam_error}) when various terminals of the drive system are connected to the terminal 2 of the response system one by one.  First we study the dynamics of the circuit when the terminal 2 of the drive systeme is connected to the terminal 2 of the response system, compactly represented as (2,2).   In this configuration, the drive system variable $v_2$ is connected to the response system variable $v_2'$ through a buffer and coupling resistor $R_c$.  This is obviously the case of two identical systems connected by an error feedback coupling. The state equation of this coupled circuit is obtained (Fig.~\ref{chuam_error}) using Kirchoff's laws.
The governing circuit equation is given by\\
Drive system : 
\begin{subequations}
\begin{eqnarray} 
C_1\frac{dv_1}{dt} &=& (1/R_1)(v_{LC}-v_1) + (1/R_2)(v_2-v_1), \\
C_2\frac{dv_2}{dt} &=& (1/R_2)(v_1-v_2) + (1/R_3)(v_3-v_2), \\
C_3\frac{dv_3}{dt} &=& (1/R_3)(v_2-v_3) - i_N, \\
C_L\frac{dv_{CL}}{dt} &=& (1/R_1)(v_1-v_{LC}) + i_L,  \\
L\frac{di_L}{dt} &=&-v_{LC}.
\end{eqnarray}
\label{cop_errorcir_eqn1} 
\end{subequations}
Response system : 
\begin{subequations}
\begin{eqnarray} 
C_1'\frac{dv_1'}{dt} &=& (1/R_1')(v_{LC}'-v_1') + (1/R_2')(v_2'-v_1'), \\
C_2'\frac{dv_2'}{dt} &=& (1/R_2')(v_1'-v_2') + (1/R_3')(v_3'-v_2') + \epsilon'(v_2-v_2'), \\
C_3'\frac{dv_3'}{dt} &=& (1/R_3')(v_2'-v_3') - i_N', \\
C_L'\frac{dv_{CL}'}{dt} &=& (1/R_1')(v_1'-v_{LC}') + i_L',  \\
L'\frac{di_L'}{dt} &=&-v_{LC}'.
\end{eqnarray} 
\label{cop_errorcir_eqn2}
\end{subequations}
The $v - i$ characteristics of the  Chua's diode is given by
\begin{eqnarray}
i_N=g(v) &=& G_b v+0.5(G_a-G_b) [|v+B_p|-|v-B_p|], 
\label{vi_char1}
\end{eqnarray}
where $G_a$ and $G_b$ are the inner and outer slopes of the characteristic curve
respectively.   Here $\pm B_p$ denote the break point of the characteristic
curve.  The values of the circuit  elements are fixed at $L=18.0~mH$,
$C_1=C_2=C_3=4.7~nF$, $C_L=100.0~nF$, $R_1=R_2=R_3=610~\Omega$, $L'=18.0~mH$, $C_1'=C_2'=C_3'=4.7~nF$,$C_L'=100.0~nF$ and $R_1'=R_2'=R_3'=610~\Omega$.  Now Eqns.~(\ref{cop_errorcir_eqn1}) and (\ref{cop_errorcir_eqn2})  can be rescaled as follows, $v_1=x B_p$, $v_2=y B_p$, $v_3=z B_p$, $v_{LC}=u B_p$, $i_L=(B_p G)h$, $t=C_L \tau /G$, $G=1/R_3$, $v_1'=x' B_p$, $v_2'=y' B_p$, $v_3'=z' B_p$, $v_{LC}'=u' B_p$, $i_L'=(B_p G)h'$ and $\epsilon'=1/R_c$.  We then redefine $\tau$ as $t$.  The rescaled version of the state equations is given as follows. \\
Drive system : 
\begin{subequations}
\begin{eqnarray} 
\dot{x} &=& \beta_1(u-x)+\beta_2(y-x), \;\;\; (\cdot = d/dt) \\
\dot{y} &=& \beta_3(x-y)+\beta_4(z-y), \\
\dot{z} &=& \alpha[(y-z)-g(z)], \\
\dot{u} &=& \gamma (x-u)+h, \\
\dot{h} &=& -\beta_0 u,
\end{eqnarray}
\label{cop_errornor_eqn1}
\end{subequations}
Response system : 
\begin{subequations}
\begin{eqnarray} 
\dot{x'} &=& \beta_1'(u'-x')+\beta_2'(y'-x'), \\
\dot{y'} &=& \beta_3'(x'-y')+\beta_4'(z'-y')+\epsilon(y-y'), \\
\dot{z'} &=& \alpha'[(y'-z')-g(z')], \\
\dot{u'} &=& \gamma' (x'-u')+h', \\
\dot{h'} &=& -\beta_0' u',
\end{eqnarray}
\label{cop_errornor_eqn2}
\end{subequations}
where $\alpha = (C_L/C_3R_3G)$, $\gamma=1/(GR_1)$, $\beta_0 = C_L/(LG^2)$, $\beta_1 = C_L/(C_1GR_1)$, $\beta_2 = C_L/(C_1GR_2)$, $\beta_3 = C_L/(C_2GR_2)$, $\beta_4 = C_L/(C_2GR_3)$, $\alpha' = (C_L'/C_3'R_3'G)$, $\beta_0' = C_L'/(L'G^2)$, $\beta_1' = C_L'/(C_1'GR_1')$, $\beta_2' = C_L'/(C_1'GR_2')$, $\beta_3' = C_L'/(C_2'GR_2')$, $\beta_4' = C_L'/(C_2'GR_3')$, $\gamma'=1/(GR_1')$ and $\epsilon = \epsilon' C_L'/C_2'G$.  The term $g(z)=g(z')$ is obviously represented in the rescaled form as 
\begin{equation}
g(z) = bz + 0.5(a - b) [|z + 1| - |z - 1|].  
\label{vi_char2}
\end{equation}
Here, $a=G_a/G=-0.4826$ and $b=G_b/G=-0.26035$. The dynamics underlying Eqs.~(\ref{cop_errornor_eqn1}) and (\ref{cop_errornor_eqn2}) now depends on the rescaled parameters, which are fixed due to the present experimental set up as $\alpha=\alpha'=\beta_1=\beta_2=\beta_3=\beta_4=\beta_1'=\beta_2'=\beta_3'=\beta_4'=21.2765$, $\gamma=\gamma'=1.0$, $\beta_0=\beta_0'=2.0672$ and $\epsilon=1.3$.
\begin{figure}
\centering
\includegraphics[width=0.6\columnwidth]{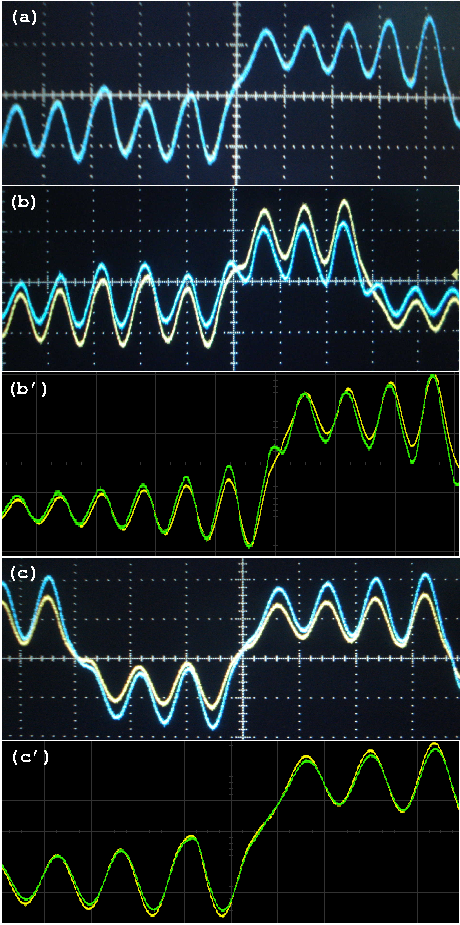}
\caption{\label{2_t_all} Experimentally observed drive (yellow, $v_2$) and response (blue, $v_2'$) wave forms showing (a) complete synchronization for (2,2) coupling, (b)  anticipatory synchronization for (1,2) coupling, (b') anticipatory synchronization for (1,2) with the response signal amplified by a factor $1.47$ (yellow, $v_2$) and response (green, $v_2'$), (c) lag synchronization (3,2) coupling and (c') lag synchronization for (3,2) with the response signal amplified by a factor $0.7$ (yellow, $v_2$) and response (green, $v_2'$). Vertical scale $2$v/div.: horizontal scale $400\mu s$/div for (a), (b) and (c), horizontal scale $260\mu s$/div for (b') and (c').}
\end{figure}

\begin{figure}
\centering
\includegraphics[width=0.55\columnwidth]{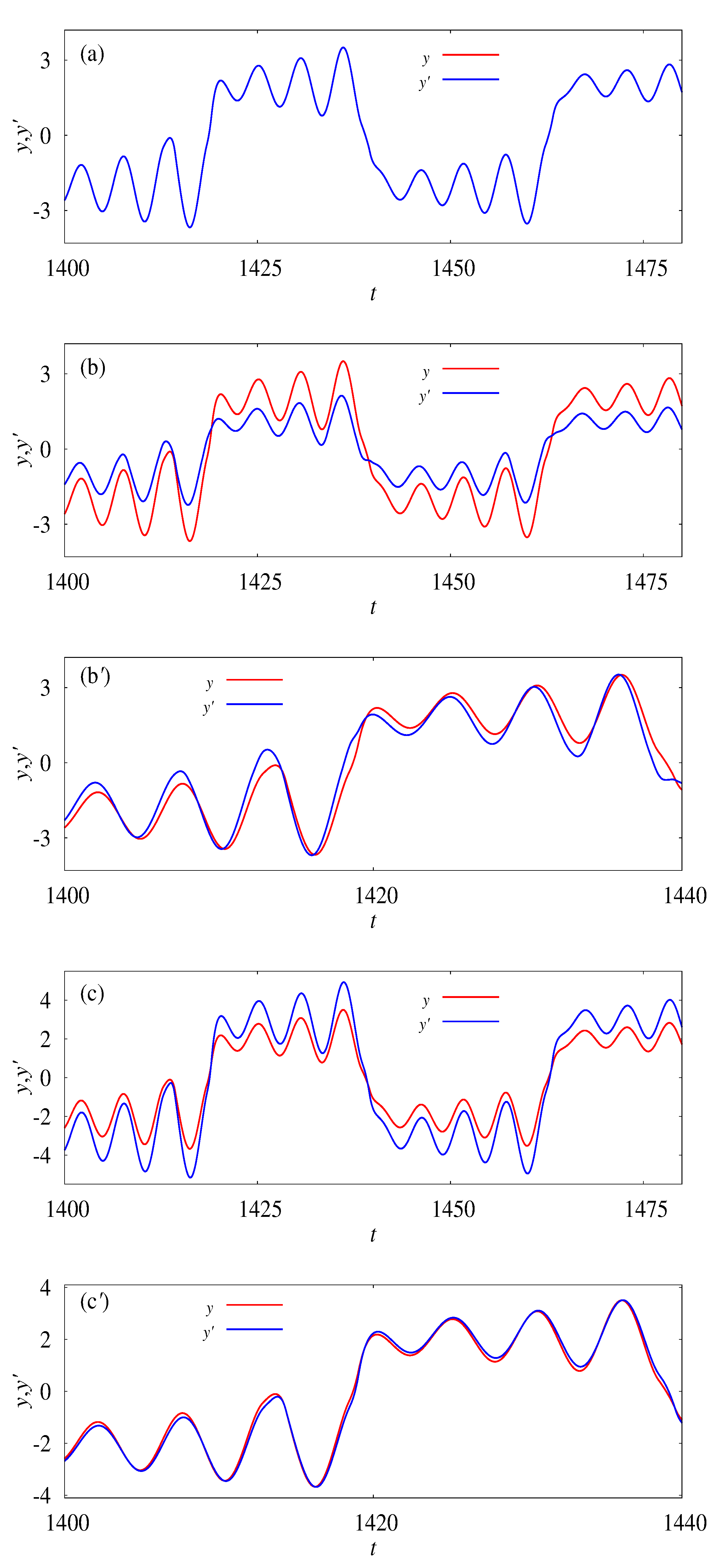}
\caption{\label{2_t_num} Numerically observed drive (red, $y$) and response (blue, $y'$) wave forms showing (a) complete synchronization for (2,2) coupling, (b)  anticipatory synchronization for (1,2) coupling, (b')  anticipatory synchronization for (1,2) coupling with the response signal $y'$ amplified by a constant factor, (c) lag synchronization for (3,2) coupling and (c\')  lag synchronization for (3,2) coupling with suitably amplified response signal.}
\end{figure}

\begin{figure}
\centering
\includegraphics[width=0.8\columnwidth]{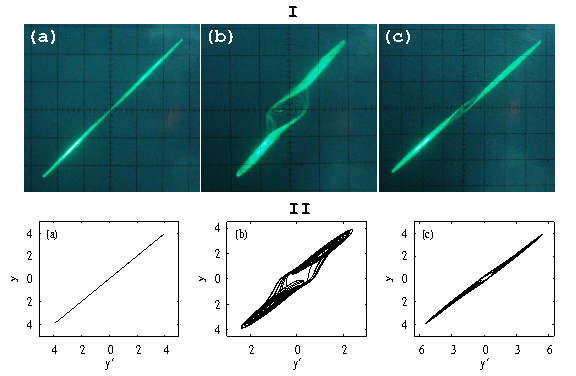}
\caption{\label{2_xy_all} I Experimentally observed drive ($v_2$) and response ($v_2'$) phase plane ($v_2-v_2'$): (a) complete synchronization for (2,2) coupling, (b)  anticipatory synchronization for (1,2) coupling and (c) lag synchronization for (3,2) coupling.  Vertical scale $1$v/div., horizontal scale $1$v/div. II Corresponding numerical results of I.}
\end{figure}

Note that the drive system drives the response by the drive component in Eqs.~(\ref{cop_errorcir_eqn2}b) or (\ref{cop_errornor_eqn2}b).  Our experimental studies and numerical analysis reveal the following. In the connection configuration $(2,2)$ (see Fig.~\ref{chuam_error}) we observe complete synchronization for $\epsilon = 1.3$.  The corresponding experimental time series of the drive and response variables ($v_2$ \& $v_2'$) are shown in Fig.~\ref{2_t_all}$(a)$ and the corresponding numerical plot is shown in Fig.~\ref{2_t_num}$(a)$.  In Fig.~\ref{2_t_all} the horizontal axis is calibrated as $400~\mu s/div.$ and the vertical axis is $2~v/div$.  From these figures, it is clear that both the phase and amplitude of the drive and response systems are the same, implying that the coupled systems of Fig.~\ref{chuam_error} exhibit complete synchronization.  The same is observed in the phase space plots as shown in Fig.~\ref{2_xy_all}$(a)$.  
\begin{figure}
\centering
\includegraphics[width=1.0\columnwidth]{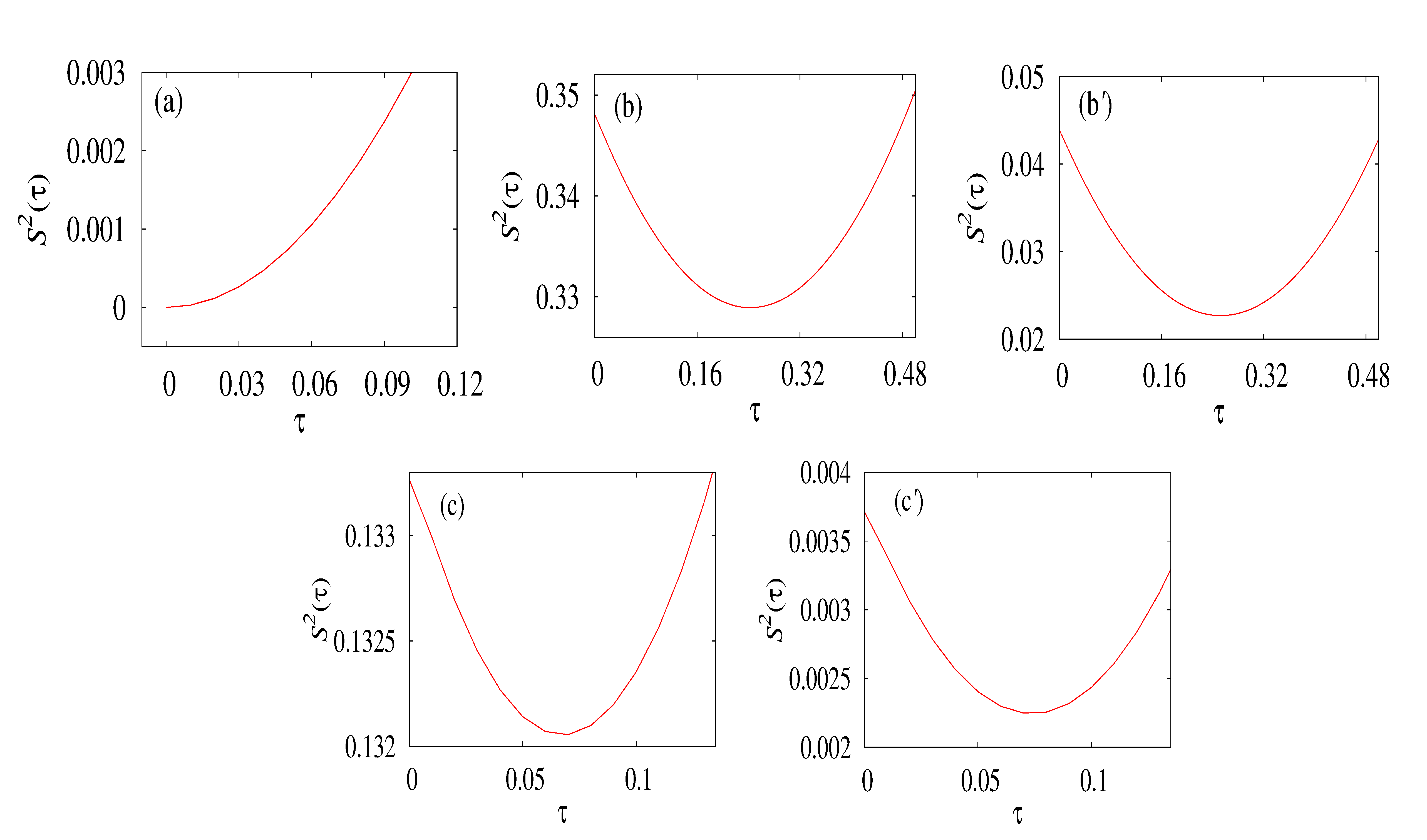}
\caption{\label{sim_2_all} Similarity function $S^2(\tau)$ corresponding to Fig.~\ref{2_t_num} confirming (a) complete synchronization for (2,2) coupling, (b)  anticipatory synchronization between unamplified drive and response signals for (1,2) coupling, (b') anticipatory synchronization between amplified drive and response signals for (1,2) coupling, (c) lag synchronization between unamplified drive and response signals for (3,2) coupling and (c') lag synchronization between unamplified drive and response signals for (3,2) coupling.}
\end{figure}
The degree of anticipatory synchronization with the corresponding time shift $\tau$ can be quantified using the similarity function~\cite{rosenblum2} defined as
\begin{equation}
S^2(\tau)=\frac{\langle [y'(t-\tau)-y(t)]^2\rangle}{[\langle y^2(t)\rangle \langle y'^{2}(t) \rangle ]^{1/2}},
\label{anti_simi_eq}
\end{equation}
where $\langle y \rangle$ is the time average of $y$.  In the case of lag synchronization the numerator of the above expression for the similarity function has to modified as $\langle [y'(t+\tau)-y(t)]^2\rangle$.  So that $\tau$ is positive for both anticipatory and lag synchronizations as per convention~\cite{dvs}.  If the signals $y(t)$ and $y^{'}(t)$ are independent, the difference between them is of the same order as the signals themselves.  If $y(t) = y^{'}(t)$, as in the case of complete synchronization, the similarity function reaches a minimum, $S^2(\tau)=0$, for $\tau=0$.  Figure \ref{sim_2_all}(a) shows the similarity function $S^2(\tau)$ as a function of the time shift $\tau$ between the drive and response systems.  One may note that the minimum of $S^2(\tau)\approx 0.00002$ occurs at $\tau \approx 0.0$ for $\epsilon=1.3$.  This indicates that there exists no time shift between the two signals in Fig.~\ref{2_t_num}(a) such that $y'(t) = y(t)$ demonstrating complete synchronization. 

Let us now study the connection configuration $(1,2)$, that is, the drive variable $v_1$ is coupled to
the response system variable $v_2'$ and correspondingly Eq.~(\ref{cop_errorcir_eqn2}b) and
Eq.~(\ref{cop_errornor_eqn2}b) get modified, respectively, as
\begin{equation}
C_2'\frac{dv_2'}{dt} = (1/R_2')(v_1'-v_2') + (1/R_3')(v_3'-v_2') + \epsilon'(v_1-v_2')
\end{equation}
and
\begin{equation}
\dot{y'} = \beta_3'(x'-y')+\beta_4'(z'-y')+\epsilon(x-y'),
\end{equation}
while the other equations in (\ref{cop_errorcir_eqn2}) and (\ref{cop_errornor_eqn2}) remain the same.  
For $\epsilon=1.3$ the coupled oscillators exhibit anticipatory synchronization.  Figures~\ref{2_t_all}$(b)$ and \ref{2_t_num}$(b)$ depict the experimental and numerical time series plot of $y$ and $y'$, respectively.  From these figures we can observe that $y'$ anticipates $y$.  In other words, the response system anticipates the drive system, thereby we can infer the anticipatory synchronization of the system shown in Fig.~\ref{chuam_error}.  The corresponding phase space plot is shown in Fig.~\ref{2_xy_all}$(b)$.  We can again use the similarity function $S^2(\tau)$ to characterize anticipatory synchronization.  For the case of a nonzero value of time shift $\tau$, if $S^2(\tau) \approx 0$, there exists a time shift $\tau$ between the two signals $y(t)$ and $y^{'}(t)$ such that $y^{'}(t-\tau)=y(t)$, demonstrating anticipatory synchronization.  Figure \ref{sim_2_all}(b) shows the similarity function $S^2(\tau)$ as a function of the time shift $\tau$.  One may note that the minimum of $S^2(\tau)\approx 0.328$ (see Fig.~\ref{sim_2_all}(b)) occurs at $\tau \approx 0.24$ for $\epsilon=1.3$.  This indicates that the response system leads the drive system by a time shift $\tau \approx 0.24$ time units. However, we note that $S^2(\tau)$ is not close to zero, and this is because the amplitude of oscillations of the response and drive systems are mismatched.  By suitably choosing an amplification factor, one can match the amplitudes of both the systems and make $S^2(\tau)\approx 0$.  For example by an amplification of the response signal $y'$ by a factor  $1.73$, we find the similarity function $S^2(\tau)\approx 0$ for the time shift $\tau=0.24$ (see Figure ~\ref{sim_2_all}(b')). In Figs.~\ref{2_t_num}(b') and \ref{2_t_all}(b') the amplified response signal $y'$ and drive signal $y$ are plotted against time.

Let us now study the dynamics of the coupling configuration $(3,2)$ for which the Eq.~\ref{cop_errorcir_eqn2}(b) and Eq.~\ref{cop_errornor_eqn2}(b) get modified as
\begin{equation}
C_2'\frac{dv_2'}{dt} = (1/R_2')(v_1'-v_2') + (1/R_3')(v_3'-v_2') + \epsilon'(v_3-v_2')
\end{equation}
and
\begin{equation}
\dot{y'} = \beta_3'(x'-y')+\beta_4'(z'-y')+\epsilon(z-y').
\end{equation}
For $\epsilon=1.3$ and the $z$-drive component coupled through one way coupling to the response subsystem, the coupled oscillators exhibit lag synchronization.  Experimental and numerical plots of time series of $v_{2}$ and $v_{2}'$ or $y$ and $y'$ are shown in Figs.~\ref{2_t_all}$(c)$ and \ref{2_t_num}$(c)$, respectively.  In this case, the response system variable $y'$ lags the drive system variable $y$.  In other words the circuit (Fig.~\ref{chuam_error}) exhibits lag synchronization.  The phase space plots for lag synchronizations is shown in Fig.~\ref{2_xy_all}$(c)$.  The similarity function $S^2(\tau)$ characterizes the lag synchronization with a positive time shift $\tau$ instead of the negative time shift $\tau$ in Eq.~(\ref{anti_simi_eq}).  The curve in Fig.~\ref{sim_2_all}(c) shows the similarity function $S^2(\tau)$ vs $\tau$. The minimum of similarity function becomes $S^2(\tau)\approx 0.13$ at $\tau \approx 0.07$ indicating that there is a time shift [Fig.~\ref{2_t_num}(c)] between the drive and response signals $y(t)$ and $y^{'}(t)$, such that $y^{'}(t+\tau)\approx y(t)$, confirming the occurrence of lag synchronization.  The minimum of $S^2(\tau)\approx 0.13$ corresponds to a lag synchronization with lag time $\tau \approx 0.07$ between $y(t)$ and $y^{'}(t)$ in Fig.~\ref{2_t_num}$(c)$.  Note here that the minimum of $S^2(\tau$) is again not close to zero because of the mismatch in the amplitude of the drive and response signals (even though the phases are matched).  With suitable amplification of the response signal (by a factor $0.7$, see in Figs.~\ref{2_t_all}$(c')$ and \ref{2_t_num}$(c')$), we find the minimum of $S^2(\tau)$ to be approximately zero, which is shown in Fig.~\ref{sim_2_all}(c').

We wish to note here that the lag/anticipatory synchronization exhibited by the system is only an average lag/average anticipatory synchronization and it is because the output of the RC network depends on the frequency of the input signal.  We have used 20,000 data points with a step size of 0.01 for calculating all the similarity function after leaving a 
transient of 1,50,000 data points.  We wish to emphasize that the value of $\tau$ calculated  using the similarity function is only an average value and we have used a fairly large set of data points in calculating it.  We have also tried using a larger data points in calculating similarity function but
it is found that it does not affect the value of $\tau$.   We also wish to point out that the amplitude
difference between the drive and response signal is also approximately a constant because the output signal of RC network depends on the frequency of the input signal.  Therefore even after suitable amplification of
the response signal, there will be a slight mismatch in the amplitude with the drive signal.  
\subsection{Drive system connected to the terminal 1 of response system}
In this section, we study the dynamics of the two coupled circuits (Fig.~\ref{chuam_error}) when the terminal 1 of the response system is connected to one of the three terminals of the drive system.  
\begin{figure}
\centering
\includegraphics[width=0.6\columnwidth]{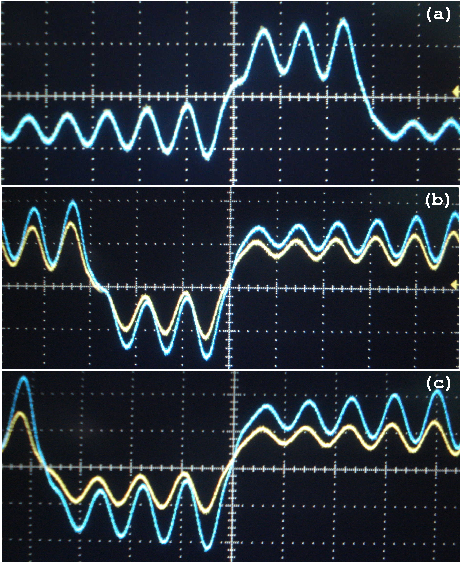}
\caption{\label{1_t_all}Experimentally observed drive (yellow, $v_2$) and response (blue, $v_2'$) wave forms showing (a) complete synchronization for (1,1) coupling, (b) lag synchronization (2,1) coupling and (c) lag synchronization (3,1) coupling. Vertical scale $2$v/div.: horizontal scale $400\mu s$/div.}
\end{figure}
\begin{figure}
\centering
\includegraphics[width=0.6\columnwidth]{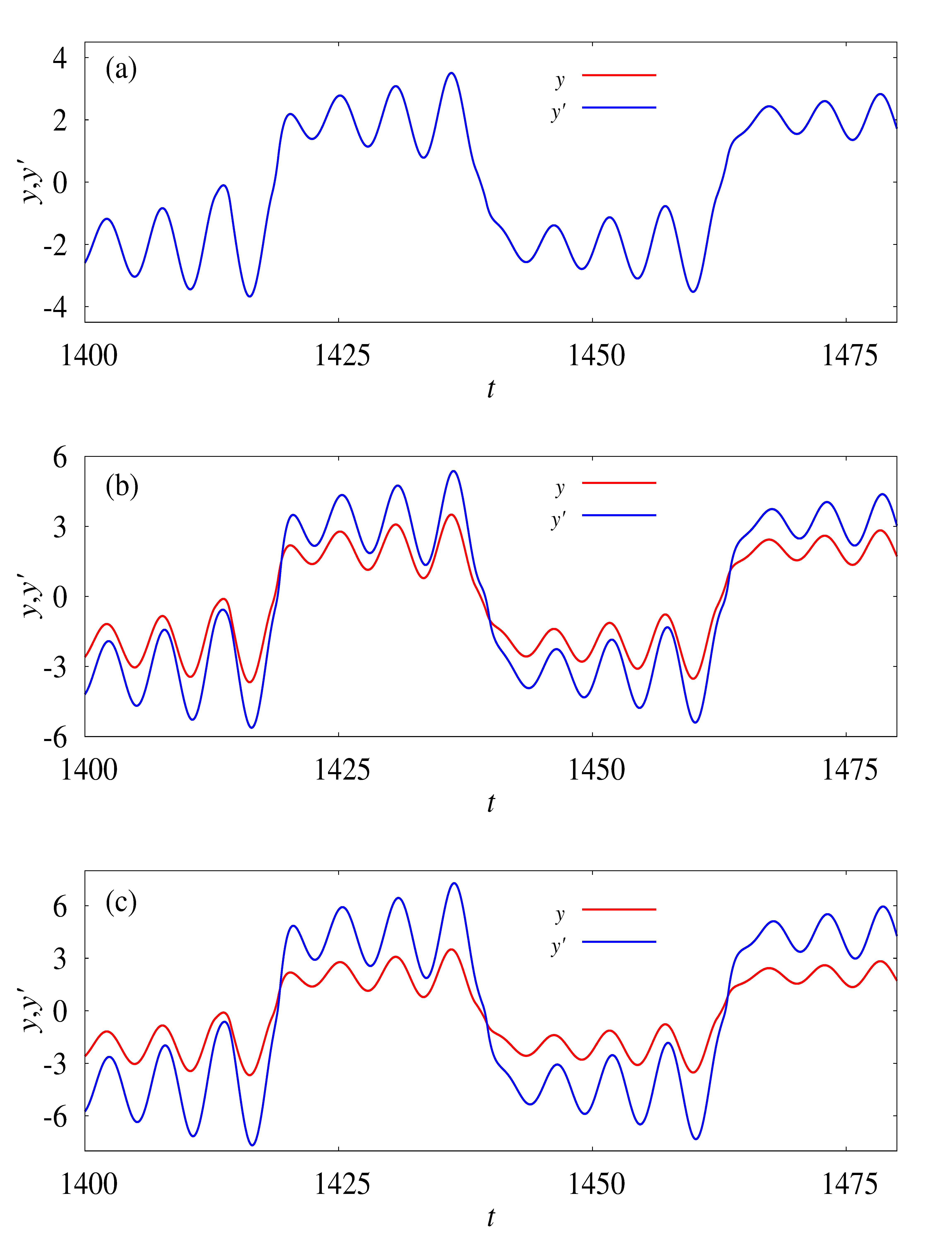}
\caption{\label{1_t_num} Numerically observed drive (red, $y$) and response (blue, $y'$) wave forms showing (a) complete synchronization for (1,1) coupling, (b) lag synchronization (2,1) coupling and (c) lag synchronization (3,1) coupling.}
\end{figure}
\begin{figure}
\centering
\includegraphics[width=1.0\columnwidth]{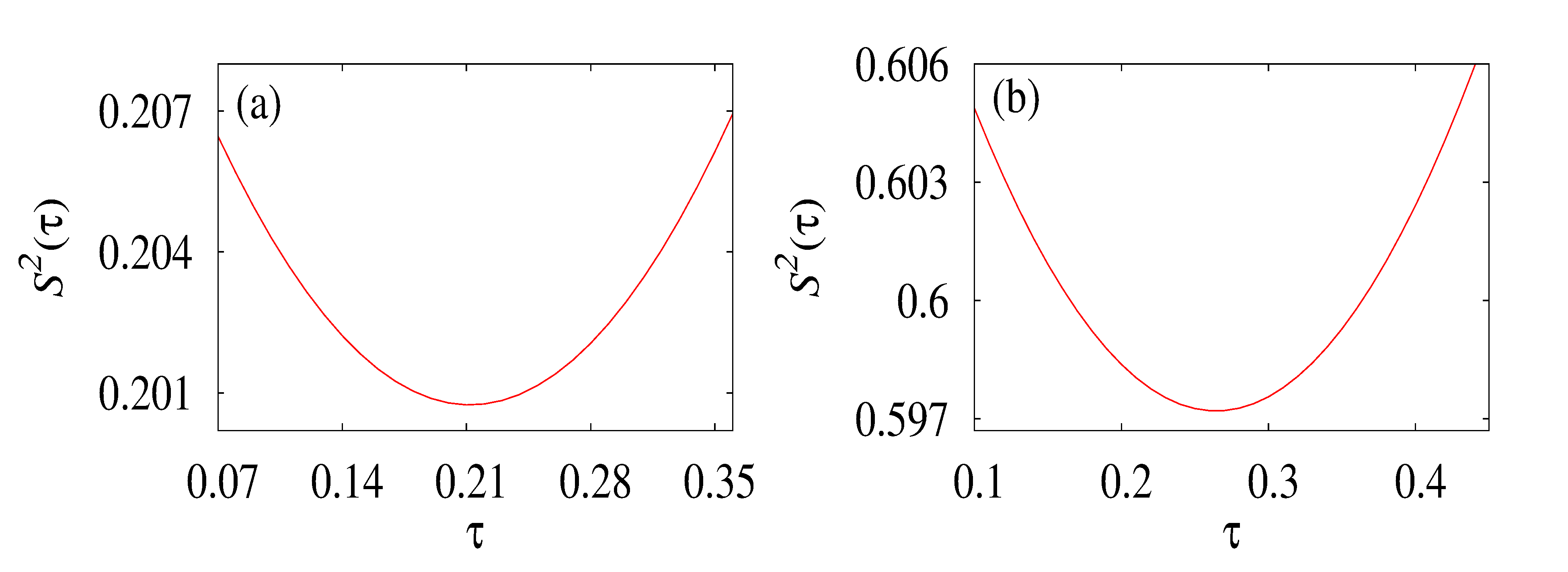}
\caption{\label{sim_1_all} Similarity function $S^2(\tau)$ corresponding to Fig.~\ref{1_t_num}(b) and \ref{1_t_num}(c) confirming (a) lag synchronization for $(2,1)$ coupling and (b) lag synchronization $(3,1)$ coupling. }
\end{figure}
\begin{figure}
\centering
\includegraphics[width=0.8\columnwidth]{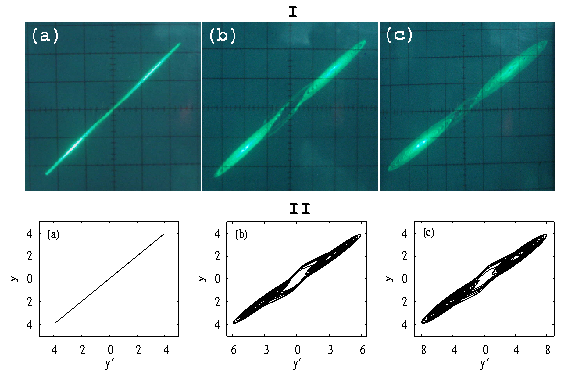}
\caption{\label{1_xy_all}I Experimentally observed drive ($v_2$) and response ($v_2'$) phase plane in ($v_2-v_2'$). (a) complete synchronization for (1,1) coupling, (b) lag synchronization (2,1) coupling and (c) lag synchronization (3,1) coupling.  Vertical scale $1$v/div., horizontal scale $1$v/div. II Corresponding numerical results of I.}
\end{figure}
For this coupling configuration, Eqs.~(\ref{cop_errorcir_eqn2}a) and (\ref{cop_errornor_eqn2}a) get modified, respectively, as
\begin{equation}
C_1'\frac{dv_1'}{dt} = (1/R_1')(v_{L'C'}-v_1') + (1/R_2')(v_2'-v_1')+ \epsilon'(V-v_1')
\end{equation}
and
\begin{equation}
\dot{x'} = \beta_1'(\omega'-x')+\beta_2'(y'-x')+\epsilon(X-x'),  
\end{equation}
while the coupling terms in equations (\ref{cop_errorcir_eqn2}b) and (\ref{cop_errornor_eqn2}b) are removed and the equations becomes
\begin{subequations}
\begin{eqnarray}
C_2'\frac{dv_2'}{dt} &=& (1/R_2')(v_1'-v_2') + (1/R_3')(v_3'-v_2'), \\
\dot{y}'&=&\beta_3'(x'-y')+\beta_4'(z'-y'),
\end{eqnarray}
\end{subequations}
where $V$ is any one of the drive signals $v_1$, $v_2$ and $v_3$ and $X$ is either $x$, $y$ or $z$.  
 
As in the previous subsection we first study the dynamics of the coupled system when terminal $1$ of the response system is connected to the terminal $1$ of the drive system. In other words, when the response circuit voltage $v_1'$ is connected with the drive circuit voltage $v_1$ through a coupling resistor $R_c$, we observe complete synchronization. Time series of the drive ($v_2$) and response ($v_2'$) voltage signals are shown in Fig.~\ref{1_t_all}$(a)$ and the corresponding numerical plot is shown in Fig.~\ref{1_t_num}$(a)$.  The same is also observed in the phase space plots, shown in Fig.~\ref{1_xy_all}$(a)$.  

Next we study the coupling configuation (2,1). For this configuration we find that the drive and the response systems are in lag synchronization for $\epsilon=1.3$.  Figures~\ref{1_t_all}$(b)$ and \ref{1_t_num}$(b)$ depict the time series plot of $y$ and $y'$ numerically and experimentally.  From these figures we observe that $y'$ lags with $y$.  
The corresponding phase space plots are shown in Fig.~\ref{1_xy_all}$(b)$.  Figure~\ref{sim_1_all}(a) shows the similarity function confirming lag synchronization.  

Finally we study the system for the coupling configuration $(3,1)$. This coupled circuit exhibits lag synchronization again.  The time series plots (experimental and numerical) showing the presence of lag synchronization are given in Figs.~\ref{1_t_all}(c) and \ref{1_t_num}(c), respectively.  The phase plane plots are shown in Fig.~\ref{1_xy_all}$(c)$.  The lag synchronization is confirmed by the minimum of the similarity function (Fig.~\ref{sim_1_all}(b)) and the amount of lag in the $(3,1)$ coupling configuration is found to be higher than the lag  in $(2,1)$ configuration.  From Figs.~\ref{sim_1_all}(a,b) we find that the similarity function $S^2(\tau)$ is minimum at $\tau=0.21$ for $(2,1)$ coupling and $S^2(\tau)$ is minimum at $\tau=0.27$ for the coupling configuration $(3,1)$.  Again we find that the minimum values of the similarity function $S^2(\tau)$ for both the configurations (2,1) and (3,1) are not close to zero and the reason for this is the amplitude mismatch explained in Section 2.1.  With suitable amplification to the response system the minimum of $S^2(\tau)$ becomes quite close to zero in each of the cases.
      
\subsection{Drive system connected to the terminal 3 of the response system}
In this subsection, we study the nature of synchronization in the coupled circuits (Fig.~\ref{chuam_error}) in the configuration $(m,3)$, $m=1,2,3$.  In this configuration, the response system equations, Eqs.~(\ref{cop_errorcir_eqn2}c) and (\ref{cop_errornor_eqn2}c), get modified as
\begin{equation}
C_3'\frac{dv_3'}{dt} = (1/R_3')(v_2'-v_3') - i_N'+ \epsilon'(V-v_3')
\end{equation}
and
\begin{equation}
\dot{z'} = \alpha'[(y'-z')-g(z')]+\epsilon(Z-z').  
\end{equation}
where $V$ denotes any one of the drive signals $v_1$, $v_2$ and $v_3$ and $Z$ denotes any one of the scaled variables $x$, $y$ and $z$.  
For this coupling configuration the coupling terms in equations (\ref{cop_errorcir_eqn2}b) and (\ref{cop_errornor_eqn2}b) are removed.

\begin{figure}
\centering
\includegraphics[width=0.6\columnwidth]{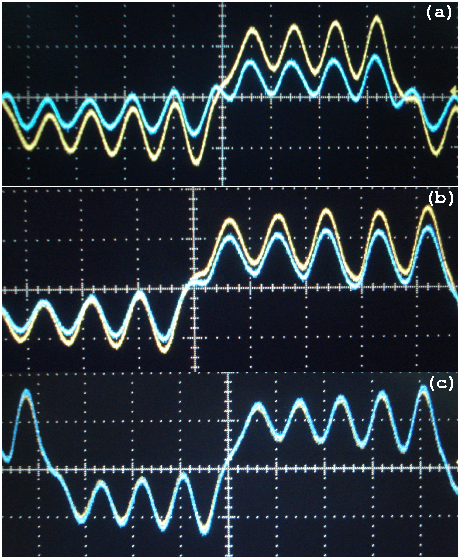}
\caption{\label{3_t_all} Experimentally observed drive (yellow, $v_2$) and response (blue, $v_2'$) wave forms showing (a) anticipatory synchronization for (1,3) coupling, (b) anticipatory synchronization (2,3) coupling and (c) complete synchronization (3,3) coupling.  Vertical scale $2$v/div.: horizontal scale $400 \mu s$/div.}
\end{figure}

\begin{figure}
\centering
\includegraphics[width=0.6\columnwidth]{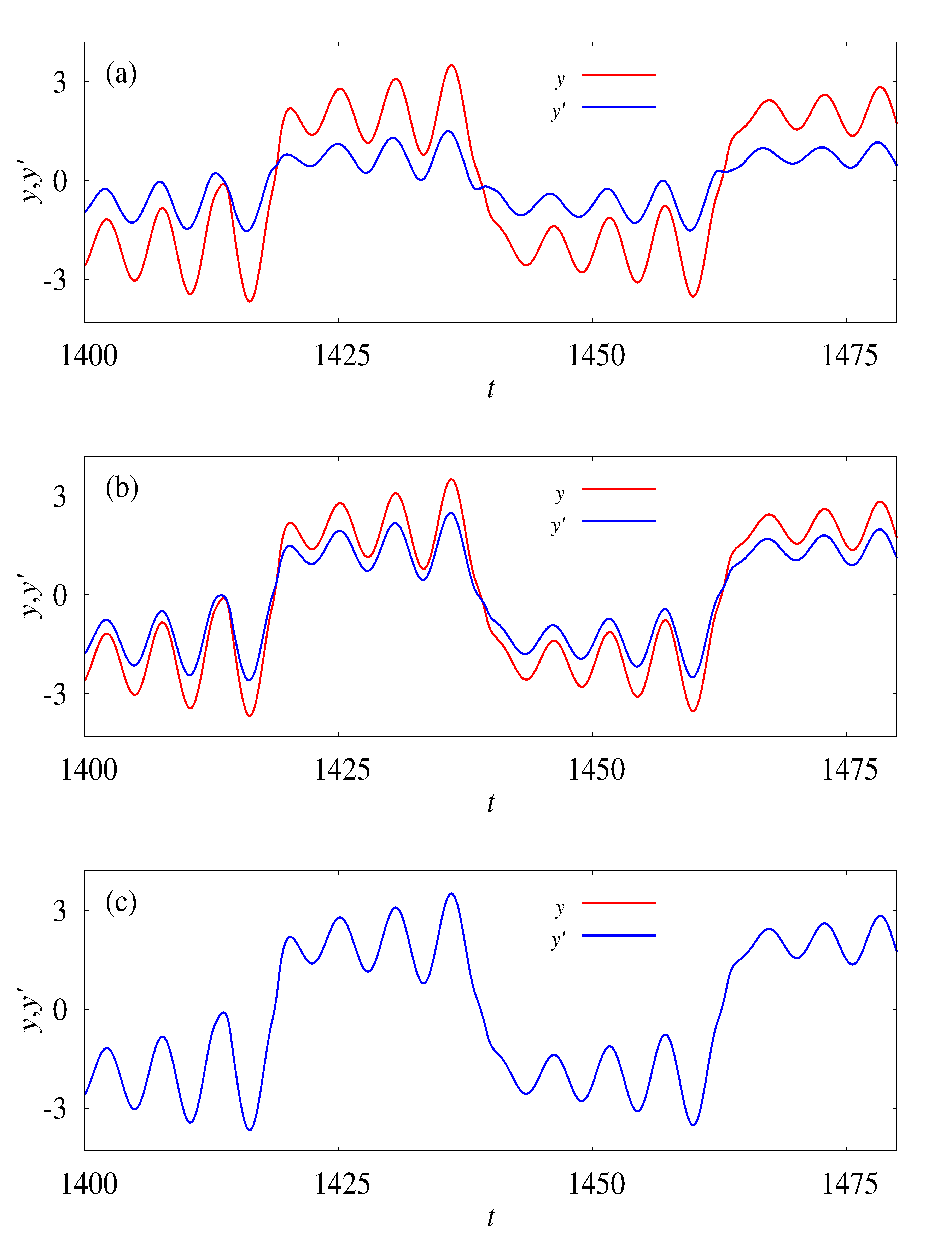}
\caption{\label{3_t_num} Numerically observed drive (red, $y$) and response (blue, $y'$) wave forms showing (a) anticipatory synchronization for (1,3) coupling, (b) anticipatory synchronization (2,3) coupling and (c) complete synchronization (3,3) coupling.}
\end{figure}

\begin{figure}
\centering
\includegraphics[width=1.0\columnwidth]{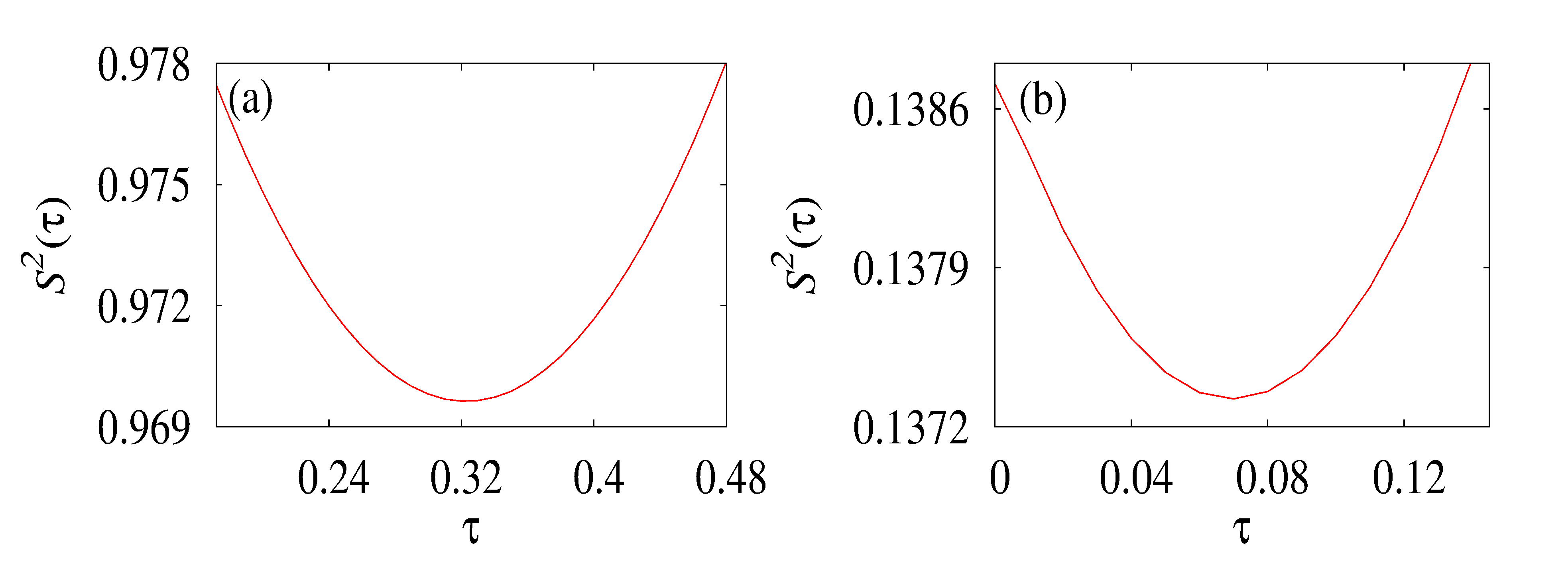}
\caption{\label{sim_3_all} Similarity function $S^2(\tau)$ corresponding to Fig.~\ref{3_t_num}(a) and \ref{3_t_num}(b) anticipatory synchronizations.  }
\end{figure}

\begin{figure}
\centering
\includegraphics[width=0.8\columnwidth]{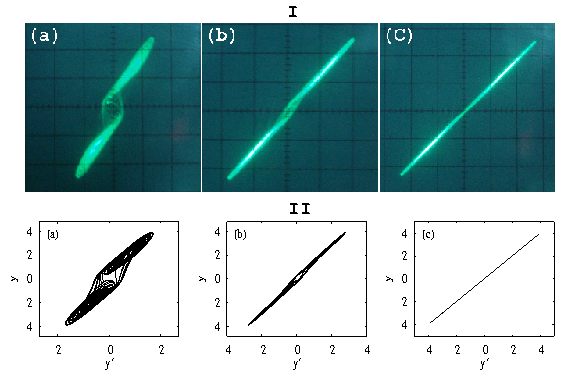}
\caption{\label{3_xy_all}I Experimentally observed drive ($v_2$) and response ($v_2'$) phase plane in ($v_2-v_2'$). (a) anticipatory synchronization for (1,3) coupling, (b) anticipatory synchronization (2,3) coupling and (c) complete synchronization (3,3) coupling.  Vertical scale $1$v/div., horizontal scale $1$v/div. II Corresponding numerical results of I.}
\end{figure}

Let us first consider the coupling configuration $(1,3)$. This coupling configuration exhibits anticipatory synchronization.  The time series plots of drive and response system are shown in Figs.~\ref{3_t_all}(a)(experimental) and \ref{3_t_num}(a) (numerical), respectively.  Figure~\ref{3_xy_all} depicts the corresponding phase plane plots of the drive and response systems.  These figures clearly indicate the anticipatory synchronization exhibited by the coupled circuits.  Again the synchronization is characterized using similarity function which is plotted in Fig.~\ref{sim_3_all}(a).  The minimum of the similarity function is found at $\tau=0.32$.  

Changing the coupling configuration to $(2,3)$ also results in anticipatory synchronization.  The time series plots obtained experimentally and numerically are shown in Figs.~\ref{3_t_all}(b) and \ref{3_t_num}(b) respectively.  The corresponding similarity function given in Fig.~\ref{sim_3_all}(b) clearly shows the anticipatory synchronization of the coupled system ($S^2(\tau)$ minimum at $\tau=0.07$)].  As mentioned in the previous cases, with suitable amplification to the response signal one can make the minimum of $S^2(\tau)$ close to zero.   From the similarity function plot we find that the amount of lead in the configuration $(2,3)$ is lower than the lead in the $(1,3)$ configuration.   Figure~\ref{3_xy_all}(b) depicts the corresponding phase plane plots of drive and response systems.   

Finally, we connect the drive and response systems in the configuration $(3,3)$ and in this configuration the coupled systems exhibit identical/complete synchronization.  Figures.~\ref{3_t_all}(c), \ref{3_t_num}(c) and \ref{3_xy_all}(c) clearly show the existence of complete synchronization.    

A summary of the nature of synchronizations is provided in detail in Table~\ref{tab}.  
\begin{table}
\caption{Summary of the nature of synchronizations in the two coupled circuit systems under different couplings}
\begin{center}
\label{tab}
  \begin{tabular}{|c|c|l|}
    \hline
    Drive   & Response & Nature of synchronization\\
    \hline
    1 & 1 & Complete \\
    2 & 1 & Lag \\
    3 & 1 & Lag \\

    1 & 2 & Anticipatory  \\
    2 & 2 & Complete  \\
    3 & 2 & Lag \\
    
    1 & 3 & Anticipatory \\
    2 & 3 & Anticipatory \\
    3 & 3 & Complete \\
    
    \hline
  \end{tabular}
\end{center}
\end{table}

\section{Summary}
In this paper we have developed a novel procedure by which lag, anticipatory and complete sychronizations are  observed in a coupled system without using explicit time delay or parameter mismatch.  We have experimentally observed these synchronizations in a RC phase shifted network of modified Chua's ciruit and confirmed the results by numerical studies for different coupling configurations.  Anticipatory, complete and lag synchronizations are observed when the second terminal of the response system is connected to the first, second and third terminals,  respectively, of the drive system.  With a simple switching of the connecting terminals of the drive and the response systems we observed different types of chaotic synchronization.   The observed anticipatory, complete and lag synchronizations are characterized by similarity functions. 

\section*{Acknowledgments}
K. Srinivasan acknowledges SERB-DST Fast Track scheme for Young Scientists for support (SR/FTP/PS-117/2013). V.K. Chandrasekar acknowledges the financial support of an Indian National Science Academy Young Scientist project.
M. Lakshmanan has been supported by a DAE Raja Ramanna Fellowship.  
%
\bibliographystyle{model1a-num-names}
\bibliography{<your-bib-database>}

\end{document}